\documentclass[twocolumn,showpacs,preprintnumbers,amsmath,amssymb,prl]{revtex4}
\usepackage{epsf,amsmath,amssymb,verbatim,color,multirow,pifont}
\usepackage{graphicx}

\begin{document}
\title{Percolation Transitions in Scale-Free Networks under Achlioptas Process}

\author{Y.S. Cho$^1$, J.S. Kim$^1$, J. Park$^1$, B. Kahng$^1$, and D. Kim$^{1,2}$}
\affiliation{{$^1$ Center for Theoretical Physics and Department
of Physics and Astronomy, Seoul National University, Seoul
151-747, Korea}\\
{$^2$ School of Physics, Korea Institute for Advanced Study, Seoul
130-722, Korea}}
\date{\today}

\begin{abstract}
It has been recently shown that the percolation transition is
discontinuous in Erd\H{o}s-R\'enyi networks and square lattices in
two dimensions under the Achlioptas Process (AP). Here, we show
that when the structure is highly heterogeneous as in scale-free
networks, a discontinuous transition does not always occur: a
continuous transition is also possible depending on the degree
distribution of the scale-free network. This originates from the
competition between the AP that discourages the formation of a
giant component and the existence of hubs that encourages it. We
also estimate the value of the characteristic degree exponent that
separates the two transition types.
\end{abstract}

 \maketitle

The Achlioptas process (AP) is a network evolution process in
which the number of vertices is fixed as $N$, and edges are added
one by one at each time step following a given rule that prevents
the formation of a target pattern. Recently, Achlioptas et
al.~\cite{ap} studied the percolation transition (PT) for the
Erd\H{o}s-R\'enyi (ER) model~\cite{er} following an AP rule,
called the product rule (PR) in which the formation of a giant
component is discouraged. In their study, the network was
developed by choosing between one of two randomly selected edges;
the selected edge had a lower value of the product of the size of
the two components that edge is joining. They found that the giant
component emerged suddenly at a percolation threshold $p_c$, and
that the PT was first-order. This transition pattern differs
drastically from the continuous PT occurring in the conventional
ER model. The transition is delayed as $p_c \equiv L_c/N\approx
0.88$, larger than $p_c=1/2$ for the conventional ER model, where
$L_c$ is the number of edges added to the system up to the
transition point. More recently, Ziff~\cite{ziff} found the same
first-order transition in the two-dimensional bond percolation
clusters under AP. Similar explosive transition pattern has also
been observed in a jamming transition model of Internet
packets~\cite{moreno}.

Here, we study the PT in a model scale-free (SF) network under the
AP rule. SF networks contain heterogeneous degrees, and their
distribution follows a power law, $P_d(k)\sim k^{-\lambda}$. To
construct artificial SF networks, a stochastic model called the
Chung and Lu (CL) model~\cite{chung} is used. Similar to the ER
model and the static model~\cite{static}, the CL model starts with
a fixed number of $N$ vertices indexed $i=1,\dots,N$. Then a
vertex $i$ is assigned a weight of $w_i=(i+i_0-1)^{-\mu}$, where
$\mu \in [0,1)$ is a control parameter, and $i_0 \propto 
N^{1-1/2\mu}$~\cite{dkim_note}
for $1/2 < \mu < 1$ and $i_0=1$ for $\mu < 1/2$.
Then, two different vertices ($i,j$) are selected with their probabilities
equal to the normalized weights, $w_i/\sum_k w_k$ and $w_j/\sum_k
w_k$, respectively, and an edge is added between them unless one
already exists. This process is repeated until $pN$ edges are
created in the system. The obtained network is SF in degree
distribution with the exponent $\lambda=1+1/\mu$. Henceforth, we
will use the CL model to study the percolation transition of scale
free networks in PR (SFPR).

The mechanism by which a giant component in PT forms in
conventional SF networks with $2 < \lambda < 3$ is different from
that in ER networks. In a ER network, as the number of edges
$L=pN$ increases in the system, multiple isolated small components
are created and merged together. This process continues up to the
finite percolation threshold $p_c$ where a single giant component
emerges through an abrupt coalescence of those small components.
On the contrary, in SF networks with $2 < \lambda < 3$, the
percolation threshold is zero in the thermodynamic limit. Thus,
the giant component initially develops with the largest degree
vertex as the seed, and grows continuously by aggregating
small-size components. The development and growth of the giant
component is the result of relatively high probability of a
vertex being chosen in the giant component~\cite{npb}. In the SFPR,
on the other hand, two vertex pairs are selected according to the
aforementioned weights. During the network growth, if two vertices
get selected from the same component, an edge is created between
them with no change in component size. Thus, the existence of a
giant component implies that even under AP, the probability of
growing the giant component is very high. This leads us to ask the
following question: {\sl what is the impact of introducing the AP
rule on the nature of the percolation transition in SF networks?}

\begin{figure}[h]
\includegraphics[width=0.8\linewidth]{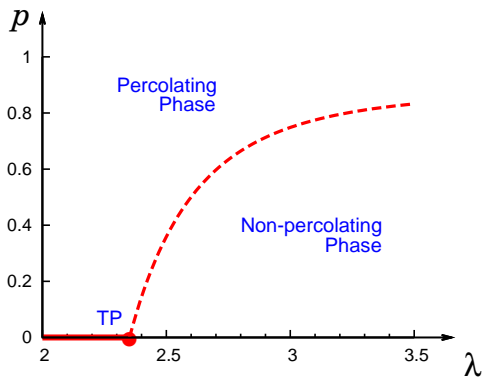}
\caption{(Color online) Phase diagram of the percolation
transition in the SFPR network. Here, $p=L/N$ is the edge
density, and $\lambda$ is the control parameter corresponding to
the degree exponent of non-PR SF networks. A second-order
(first-order) PT is represented by a solid line (dashed line). The
tricritical point is denoted as ``TP." }\label{phase_diagram}
\end{figure}

We obtain the following results by performing extensive numerical
simulations for the SFPR model: There exists a tricritical point
$\lambda_c$, estimated to be between $2.3 < \lambda_c < 2.4$, such
that when $2 < \lambda \le \lambda_c$, the transition point $p_c$
is zero in the thermodynamic limit, and the PT is second-order as
in conventional SF networks. When $\lambda > \lambda_c$, however,
$p_c$ is finite, and the transition is first-order. The jump in
the giant component size at the first-order transition point
decreases as $p_c$ decreases. The phase diagram is depicted in
Fig.~\ref{phase_diagram}. In finite-size systems, however,
$p_c(N)$ is finite even when $\lambda < \lambda_c$ and the
transition is first-order. In addition to this new feature, many
other unexpected behaviors emerge.

\begin{figure}[h]
\includegraphics[width=1.0\linewidth]{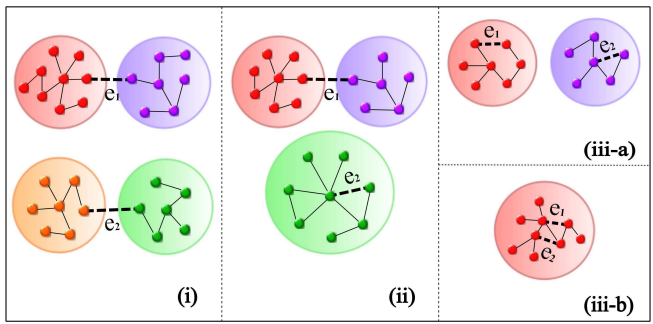}
\caption{(Color online) A schematic diagram of the selection rules
in AP for cases (i)-(iii) defined in the text. In case (i), two
inter-component edges are drawn at random, and one
of them is chosen to be connected according to the product rule
(PR). In case (ii), one edge is inter-component and the other
intra-component edge. The latter is chosen. In cases
(iii-a) and (iii-b), two intra-component edges are drawn, and one
is randomly chosen to be connected.}\label{rule}
\end{figure}

Specifically, numerical simulations are performed for the CL model
with the PR.
At each time step, two candidate edges, $e_1$ and $e_2$ are drawn
from the system with respective probabilities as described
previously and added is that that minimizes the product of the
component sizes on each end of the respective edge. Depending on
the type of the edge, there are three possible cases: (i) both
edges $e_1$ and $e_2$ are inter-component ones, (ii) one edge
$e_1$ is intra-component, and the other $e_2$ inter-component, or
(iii) both edges $e_1$ and $e_2$ are intra-component. Two subcases
of (iii) are shown in (iii-a) and (iii-b) of Fig.~\ref{rule}. For
each case, the edge added to the system is selected as follows: In
(i), the edge that minimizes the product of the component sizes on
each side of respective edge (PR) is chosen. In (ii), the edge
$e_2$ is chosen, leading to no change in component size. For
(iii), an edge is chosen randomly between the two. Schematic
picture of the selection rule in the AP is depicted in
Fig.~\ref{rule}. Henceforth $\lambda \equiv 1+1/\mu$ is a control
parameter of simulation: We find that in SFPR $\lambda$ is not the
resulting degree exponent, unlike in the conventional CL model
(see below).

\begin{figure}[h]
\includegraphics[width=1.0\linewidth]{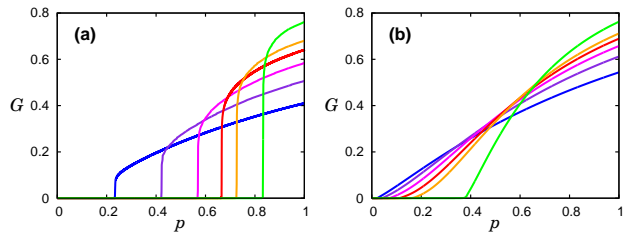}
\caption{(Color online) The fraction $G$ of the giant component
versus the edge density $p$ for (a) the CL model under AP, and (b) the
conventional CL model. Data were obtained for networks with
various control parameters $\lambda$ (2.2, 2.4,
2.6, 2.8, 3.0, and 4.0 from left (blue) to right (green)). System
size was fixed at $N=10^7$.}\label{giant}
\end{figure}

\begin{figure}[t]
\includegraphics[width=1.0\linewidth]{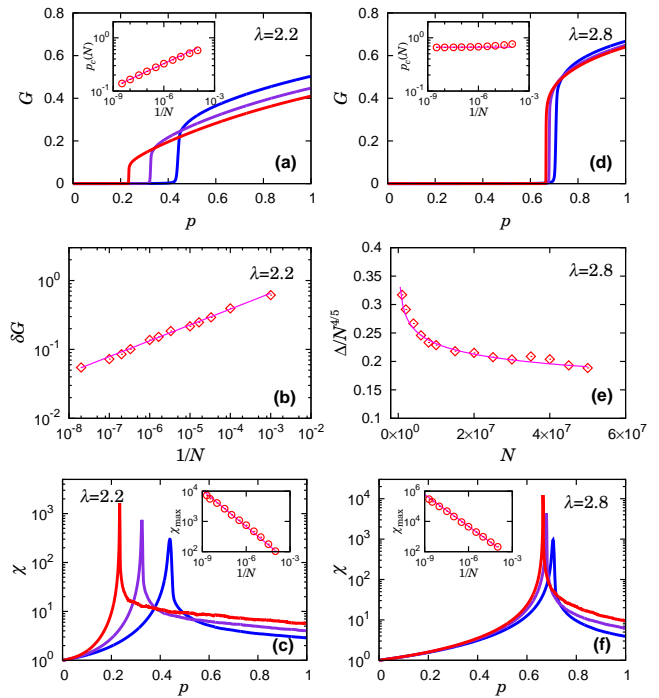}
\caption{(Color online) (a) Same plot as Fig.3 but for various
system sizes $N=10^5, 10^6$ and $10^7$, from right to left.
Control parameter $\lambda=2.2$. Inset: Plot of $p_c(N)$ versus
$1/N$. The solid line is a guideline with slope 0.15, indicating that
$p_c(\infty)\to 0$. (b) Plot of the jump $\delta G$ around
$p_c(N)$ versus $1/N$. Solid line is a guideline with slope 0.23.
(c) Susceptibility versus $p$. Inset: The peak value versus $1/N$.
(d) Same as (a) for $\lambda$=2.8. Inset: Same as the inset of
(a) for $\lambda$=2.8. Solid line is a guideline with slope 0.0,
indicating that $p_c(\infty)$ is finite. (e) Scaling plot of
$\Delta/N^{0.8}$ versus $N$ for $\lambda=2.8$, where $\Delta/N
\equiv p_1-p_0$ with $p_1$ and $p_0$ being the edge densities when
the fractions of the giant component reach $G=0.3$ and
$G=N^{-1/2}$ for the first time, respectively. (f) Same as
(c) for $\lambda=2.8$. Error bars in each data point are within
symbol sizes.}\label{finite}
\end{figure}

We measure the fraction of vertices in the giant component,
denoted as $G$, averaged over $10^2 \sim 10^4$ different network
configurations, as a function of $p$. We define the PT point,
denoted by $p_c(N)$, in a system of finite size $N$ as the point
at which the local slope of $G$ is maximal.
This position is consistent with
the peak position of the susceptibility defined below.
We also define the discontinuity of $G$~\cite{another}, denoted as $\delta G$, as
the height of the intersection point of two tangent lines, one from
the rapidly increasing transition region and the other from the
smoothly increasing curve after the jump.
Indeed, $G$ shows the first-order phase transition at $p_c(N)$
in finite size systems as shown in Fig.~\ref{giant}.
As the parameter $\lambda \to 2$ (equivalently $\mu \to 1$),
the transition point $p_c(N)$ and the
jump $\delta G$ decrease. To understand the behavior of $G(p)$ in
the $N\to \infty$ limit, numerical simulations are performed for
various system sizes in Fig.~\ref{finite}. We find that there
exists a critical value $\lambda_c$, estimated to be between $2.3
< \lambda < 2.4$, such that for $\lambda < \lambda_c$, $p_c(N)$
decreases to zero as $N$ increases (Fig.\ref{finite}(a)) in a
power-law manner $p_c(N)\sim N^{-1/\zeta}$ with $\zeta > 0$
(inset of Fig.\ref{finite}(a)), and thus $p_c(N\to \infty)\to 0$. The
exponent $\zeta$ depends on $\lambda$. For example,
$1/\zeta\approx 0.15(1)$ for $\lambda=2.2$. The jump $\delta G$ at
$p_c(N)$ decreases to zero as $\delta G \sim N^{-\beta/\zeta}$, where
the exponent $\beta$ also depends on $\lambda$.
For example, $\beta/\zeta \approx 0.23(1)$ for $\lambda=2.2$
(Fig.\ref{finite}(b)).
Thus, we conclude that the PT is continuous in the
thermodynamic limit, and Achlioptas suppression is not effective
in this case. When $\lambda > \lambda_c$, however, $p_c(N \to
\infty)$ converges to a finite value (inset of Fig.\ref{finite}(d)). The
estimated values of $p_c(N)$ for different $N$s and $p_c(\infty)$
are listed in Table~I. In finite size systems,
$p_c(N)-p_c(\infty)\sim N^{-1/\zeta}$. For example, estimated value
of the exponent $1/\zeta=0.29(1)$ for $\lambda=2.8$.
$\zeta\ne 1$ indicates that the first-order transition for $\lambda=2.8$
is not critical~\cite{note}.
To check the nature of the PT in the thermodynamic limit,
we denote $L_0=p_0N$ and $L_1=p_1N$ as the number of edges at
which the value of $G$ reaches $1/\sqrt{N}$ and 0.3, respectively. We
find that there exist a scaled quantity $\Delta/N^{0.8}$ with
$\Delta \equiv L_1-L_0$, which converges to a finite value as
$N\to \infty$ for $\lambda=2.8$ (Fig.\ref{finite}(e)). The scaling
factor $N^{0.8} < N$ indicates that the transition is of
first order~\cite{ap}. It is interesting to note that the
susceptibility, defined as $\chi\equiv \sum_s s^2 n_s$ with $n_s$,
the number of $s$-size components per node and the sum excluding
the largest component, diverges as $N\to \infty$ even when the
transition is first-order. We find that $\chi_{\rm max}\equiv
\chi(p_c(N))\sim N^{\gamma/\zeta}$ with $\gamma/\zeta \approx 0.4$
and 0.7 for $\lambda=2.2$ and 2.8, respectively, shown
in Figs.~\ref{finite}(c) and (f). Interestingly, $\gamma/\zeta\approx 0.7$
remains unchanged for $\lambda=4.0$ and $\infty$.

\begin{figure}[h]
\includegraphics[width=0.8\linewidth]{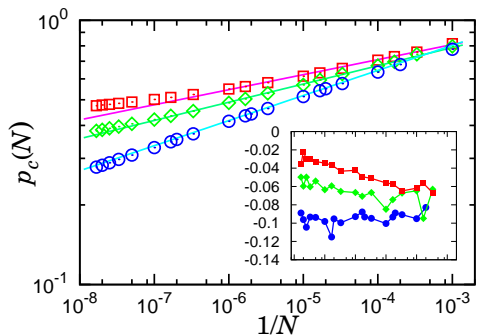}
\caption{(Color online) (a) Plot of $p_c(N)$ versus $1/N$ for
$\lambda=2.3$ ($\circ$), $2.4$ ($\diamond$), and $2.5$ ($\square$).
Error bars in each data point are within symbol sizes.
Inset: plot of successive slopes of $p_c(N)$ versus $1/N$. For
$\lambda=2.4$ ($\diamond$, green) and $2.5$ ($\square$, red), the
successive slopes approach zero, indicating that $p_c(\infty)$ is
finite. For $\lambda=2.3$ ($\circ$, blue), the successive slopes
approach a finite negative value, indicating $p_c(\infty)=0$.
}\label{critical}
\end{figure}

Since the second-order and the first-order transitions meet at
$\lambda_c$, $\lambda_c$ is a tricritical point. To estimate the
position of $\lambda_c$, we measure successive slopes of the
function $p_c(N)$ with respect to $N$ for several values of
$\lambda$ and plot them as a function of $1/N$ in the inset of
Fig.\ref{critical}. We find that the successive slopes decrease to
zero for $\lambda=2.4$ and $2.5$, while they converge to a finite
negative value for $\lambda=2.3$. Thus, we conclude that the
tricritical point is between $2.3 < \lambda_c < 2.4$, shown in
Fig.~\ref{critical}.

\begin{table}[h]
\caption{Estimated percolation threshold $p_c$ values for finite
($N_1=10^6$ and $N_2=10^7$) and infinite system sizes, and the
obtained degree exponents $\lambda^{\prime}$ at $p_c(\infty)$ for
various $\lambda$. Errors in the last decimal points are given
in parentheses.}
\begin{center}
\begin{tabular}{llllll}
\hline\hline
~~$\lambda$~~ & ~~$p_c(N_1)$~~ & ~~$p_c(N_2)$~~ & ~~$p_c(\infty)$~~
&~~ $\lambda^{\prime}(p_c)$~~ \\
\hline
~~2.2~~ & ~~0.33(1)~~~ &~~0.23(1)~~~ & ~~0~~~& ~~2.8(1)~~~\\
~~2.3~~ & ~~0.42(1)~~~ &~~0.33(1)~~~ & ~~0~~~& ~~3.0(1)~~~\\
~~2.4~~ & ~~0.49(1)~~~ &~~0.42(1)~~~ & ~~0.18(1)~~~& ~~3.1(1)~~~\\
~~2.6~~ & ~~0.60(1)~~~ &~~0.57(1)~~~ & ~~0.52(1)~~~& ~~3.5(1)~~~\\
~~2.8~~ & ~~0.68(1)~~~ &~~0.66(1)~~~ & ~~0.65(1)~~~& ~~3.8(1)~~~\\
~~3.0~~ & ~~0.73(1)~~~ &~~0.73(1)~~~ & ~~0.72(1)~~~& ~~4.2(1)~~~\\
~~4.0~~ & ~~0.83(1)~~~ &~~0.83(1)~~~ & ~~0.83(7)~~~& ~~6.3(1)~~~\\
\hline
\end{tabular}
\end{center}
\label{table1}
\end{table}

The relative frequencies of occurrence of the three cases of
(i)-(iii) of Fig.~\ref{rule} during the evolution is related to
the degree effectiveness of AP. We find that the case (i) occurs
dominantly with a probability nearly one during the period $p <
p_c(N)$, in which an attached edge connects two isolated
components, merging them into a larger component. Above $p_c(N)$,
it decays rapidly since a giant component is already there. The
cases (ii) and (iii) begin to occur when $p$ is close to $p_c$.
Next, we examine the component-size distribution during the
evolution. In early time regime $p \ll p_c(N)$, the component size
distribution exhibits an exponential decaying behavior. As $p$ is
increased, the distribution develops a hump in large-size region,
which is made through the coalescence of small-size components,
resulting in the abundance of large-size components. As $p$ just
passes $p_c$, these components finally condense into one giant
component, resulting in the disappearance of the hump and a
power-law distribution of component sizes. This behavior proves
that the self-organization process operates during the very short
transition period even under the action of AP.

\begin{figure}
\includegraphics[width=0.8\linewidth]{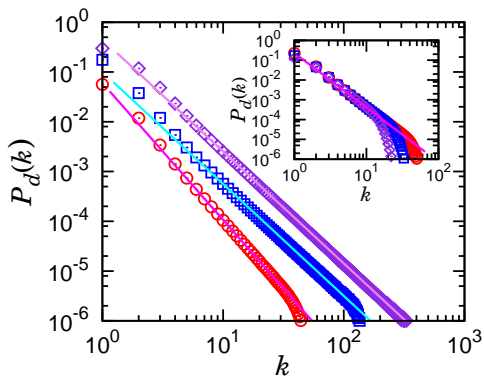}
\caption{(Color online) (a) Plot of the degree distribution
$P_d(k)$ of the SFPR network when $\lambda=2.2$. We find that the
degree exponent $\lambda^{\prime}\approx 2.8$ when $p=0.234\approx
p_c(N)$ ($\circ$,~red), and $\lambda^{\prime}\approx 2.25$ when
$p=0.8$ ($\diamond$,~purple). The degree distribution of the
conventional CL network at $p=0.234$ is drawn ($\square$,~blue)
for comparison. The system size is fixed at $N=10^7$. 
Inset: plot of the degree distributions of the SFPR network 
at $p_c(N)$ for various system sizes $N=10^5$, $10^6$, 
and $10^7$ from left (purple) to right (red), 
showing the distributions' insensitivity to system
size. Data have been shifted vertically for clear view in the main
panel, and for easier comparison in the inset. }\label{deg_dist}
\end{figure}

Unexpectedly, the tricritical point $\lambda_c$ is
located in the range $(2.3,2.4)$. To
understand the underlying mechanism, we measured the degree
distribution of the SFPR network. The degree distribution follows
a power law $P_d(k)\sim k^{-\lambda^{\prime}}$.
Yet, the exponent $\lambda^{\prime}$ varies,
depending on the edge density $p$ for a given $\lambda$, as shown
in Fig.~\ref{deg_dist}.  We find that $\lambda^{\prime}$ decreases
as $p$ increases. Numerical values of $\lambda^{\prime}$ obtained
at $p_c(N=10^7)$ as a function of $\lambda$ are listed in Table~I.
Since the degree exponent $\lambda^{\prime}$ obtained at $p_c(N)$
turns out to be insensitive to system size $N$ (the inset of
Fig~\ref{deg_dist}), $\lambda^{\prime}$ at $p_c(N=10^7)$ may be
regarded as the one at $p_c(\infty)$, even though
$\lambda^{\prime}$ is not defined at $p=0$. Interestingly, when
$\lambda < \lambda_c$, $\lambda^{\prime} \le 3$.  Thus, we can assume
that $p_c(\infty)=0$ when $\lambda^{\prime} \le 3$ at $p_c$ as
long as $\lambda < \lambda_c$. This result is reminiscent of the
well-known fact that $p_c(\infty)=0$ when $\lambda \le 3$ in
conventional uncorrelated SF networks.

In summary, we have studied the percolation transition in the
evolution of SF networks governed by AP. The nature of the phase
transition changes from continuous to discontinuous as the degree-exponent
parameter $\lambda$ is tuned past a tricritical value
$\lambda_c$ (Fig.~\ref{phase_diagram}). This phenomenon originates 
from a competition between AP that discourages the formation of 
a giant component and the existence of hubs in SF networks that encourages it.

{\sl Note added in proof:} Shortly after the submission of this
manuscript, we became aware of a similar work~\cite{santo} under
preparation. It uses a different model from ours, the
configuration model, exhibiting similar properties with some
differences.

This work is supported by KOSEF grant Acceleration Research (CNRC)
(Grant No.R17-2007-073-01001-0), and NAP of KRCF. Thank Dr.
Fortunato for sending us their paper and an anonymous
referee for introducing Ref.~\cite{critical1}.


\begin{thebibliography}{99}
\bibitem{ap} D. Achlioptas, R. M. D'Souza, and J. Spencer,
{\it Science} {\bf 323}, 1453 (2009).
\bibitem{er} P. Erd\H{o}s, A. R\'enyi, Publ. Math. Hungar. Acad. Sci. {\bf 5,} 17 (1960).
\bibitem{ziff} R. M. Ziff, Phys. Rev. Lett. {\bf 103}, 045701 (2009).
\bibitem{moreno} P. Echenique, J. Gomez-Gardenes, and Y. Moreno, Europhys. Lett. {\bf 71,} 325 (2005).
\bibitem{chung} F. Chung and L. Lu, Annals of Combinatorics {\bf 6,} 125 (2002).
\bibitem{static} K.-I. Goh, B. Kahng and D. Kim, Phys. Rev. Lett. {\bf 87,} 278701 (2001).
\bibitem{dkim_note} We use the proportionality constant $(10\sqrt{2}(1-\mu))^{1/\mu}$ 
to eliminate the degree-degree correlation.
\bibitem{npb} D.S. Lee, K.-I. Goh, B. Kahng and D. Kim, Nucl. Phys. B {\bf 696,} 351 (2004).
\bibitem{another} $\delta G$ may be measured as $\delta G_2 \equiv G(p_2)-G(p_1)$,
where $p_1$ and $p_2$ are defined through $dG/dp|_{p_1,p_2}=r dG/dp|_{p_c(N)}$ and
$p_2 > p_1$. Here, $r (< 1)$ is a tuning parameter. We find that this alternative
method does not change the $N$-dependence of $\delta G$, regardless of $r$.
\bibitem{note} The exponent $\zeta$ corresponds to $d\nu$ in the Euclidean
space, where $d$ is spatial dimension and $\nu$ is the correlation length exponent. When
the first-order transition is critical, it is known that $\nu$ becomes $1/d$ and
thus, $\zeta=1$~\cite{critical1,critical2}; however, we obtain numerically that $1/\zeta\approx 0.29,
0.69$, and $0.72$ for $\lambda=2.8, 4.0$ and $\infty$, respectively. Thus, the feature of the
first-order transition occurring in the current disordered system is different from the one
in thermal systems.
\bibitem{critical1} M. Argollo de Menezes, C.F. Moukarzel and T.J.P. Penna. Europhys.
Lett. {\bf 50}, 5 (2000).
\bibitem{critical2} M.E. Fisher and A.N. Berker, Phys. Rev. B {\bf 26}, 2507 (1982).
\bibitem{santo} F. Radicchi and S. Fortunato, arXiv:0907.0755.
\end{thebibliography}
\end{document}